\documentclass[epj]{svjour}
\usepackage{latexsym}
\usepackage{graphicx}

\begin{document}

\title{Quantum transport in honeycomb lattice ribbons with armchair 
   and zigzag edges coupled to semi-infinite linear chain leads}
\titlerunning{Quantum transport in honeycomb lattice ribbons with
   armchair and zigzag edges}
\author{Eduardo Cuansing \and Jian-Sheng Wang}
\authorrunning{E. Cuansing \and J.-S. Wang}
\institute{Department of Physics and Center for Computational Science
   and Engineering, National University of Singapore, Singapore
   117542, Republic of Singapore}
\date{Received: \today}

\abstract{
We study quantum transport in honeycomb lattice ribbons with either
armchair or zigzag edges.  The ribbons are coupled to semi-infinite
linear chains serving as the input and output leads and we use a 
tight-binding Hamiltonian with nearest-neighbor hops.  For narrow 
ribbons we find transmission gaps for both types of edges.  The center 
of the gap is at the middle of the band in ribbons with armchair edges.
This symmetry is due to a property satisfied by the matrices in the 
resulting linear problem.  In ribbons with zigzag edges the gap center 
is displaced to the right of the middle of the band.  We also find 
transmission oscillations and resonances within the transmitting 
region of the band for both types of edges.  Extending the length of 
a ribbon does not affect the width of the transmission gap, as long 
as the ribbon's length is longer than a critical value when the gap 
can form.  Increasing the width of the ribbon, however, changes the 
width of the gap.  In armchair edges the gap is not well-defined 
because of the appearance of transmission resonances while in zigzag 
edges the gap width systematically shrinks as the width of the ribbon 
is increased.  We also find only evanescent waves within the gap and 
both evanescent and propagating waves in the transmitting regions.
}
\maketitle
%
\section{Introduction \label{sec:intro}}

Graphene is a single layer of carbon atoms in a honeycomb lattice and 
serves as the basic building block of fullerenes like carbon 
nanotubes \cite{saito98}.  It has recently been found to be stable 
under ambient conditions \cite{novoselov04,novoselov05} and has unique 
charge carrier properties \cite{novoselov05b,zhang05,geim07}.  
Graphene nanoribbons are quasi-one-dimensional honeycomb lattice 
sheets with finite widths.  Because of the width termination charge 
carriers become confined and the appearance of an energy gap is 
expected \cite{son06,fujita96,nakada96,wakabayashi99,ezawa06,brey06,sasaki06,abanin06,barone06}.
For a recent review see reference \cite{castro09}.  Graphene 
nanoribbons have been fabricated using lithography \cite{han07}, 
lithography and etching \cite{chen07}, epitaxial growth and 
lithography \cite{berger06}, chemical growth \cite{li08}, and 
deposition and etching \cite{ozyilmaz07} techniques.  Measurements 
on these nanoribbons found a conductance gap that scales inversely 
with the width of the ribbons \cite{han07,ozyilmaz07}.

Possible edge types of regular graphene nanoribbons can be armchair 
or zigzag edges, on both sides of the ribbon.  Theoretical studies
using tight-binding band calculations based on the $\pi$-states of
carbon \cite{fujita96,nakada96,wakabayashi99,ezawa06} and studies on
the solution of the two-dimensional free massless Dirac equation 
using specific boundary conditions \cite{brey06,sasaki06,abanin06} 
found different transport properties for ribbons with armchair 
edges than those with zigzag edges.  Ribbons with armchair edges
can be either metallic or semiconducting depending on the width of
the ribbon while ribbons with zigzag edges are metallic.  Furthermore,
the existence of special edge states localized at the zigzag edges
gives rise to flat bands at the Fermi level in ribbons with 
zigzag edges.  On the other hand, {\it ab initio} calculations on 
graphene nanoribbons with hydrogen passivated edges found band gaps 
in both armchair and zigzag edges \cite{son06}.  Another density 
functional theory calculation, however, found that hydrogen 
termination can affect the properties of the ribbon \cite{barone06}.

In molecular electronics the search for conducting molecular wires 
that can be used in the construction of devices such as field effect 
transistors has been actively pursued.  Wires made of linked 
thiophene units \cite{otsubo01,robertson03,pearman03,bong04} and
alternating thiophene and ethynyl units \cite{tam06} have been
synthesized with lengths up to $11$ nm.  The synthesis of a chain of 
copper ions surrounded by helical ligand strands has also
recently been reported \cite{schultz08}.  Such molecular wires can 
be coupled to electrodes such as cut nanotubes \cite{guo06a,guo06b} 
or gold break junctions \cite{venkataraman06,park07}.

In this paper we study quantum transport in honeycomb lattice
ribbons using a tight-binding model with nearest-neighbor hops.
The ribbons have either armchair or zigzag edges.  In contrast to
previous theoretical and numerical studies, however, we couple 
the ribbons to semi-infinite linear chains that serve as the input 
and output leads.  Such a system may be realized experimentally as
a nanoribbon of graphene in contact with molecular wires serving 
as the leads.  We vary both the length and the width of the 
ribbons.  For a chain of hexagons with either armchair or zigzag 
edges and coupled to input and output leads we find the existence 
of a transmission gap, as long as the ribbon is longer than a 
critical length below which the gap does not form.  The center of 
this gap coincides with the middle of the band in ribbons with 
armchair edges but not with zigzag edges.  This symmetry arises 
from a property of the matrices in the resulting linear problem.  
Extending the length of the ribbon does not affect the width of 
the transmission gap.  Extending the width of the ribbon, however, 
affects the width of the gap.  In ribbons with armchair edges the 
appearance of transmission resonances obscures the exact width of
the gap while in ribbons with zigzag edges the gap systematically 
shrinks as the width of the ribbon is extended.

\section{The model \label{sec:model}}

We model the transport of a quantum particle using the tight-binding 
Hamiltonian
\begin{equation}
\mathcal{H} = \sum_{i=1}^{N_s} \epsilon_i \left| \varphi_i \right> 
     \left< \varphi_i \right| + \sum_{<ij>} v_{ij} \Bigl\{ \left| 
     \varphi_i \right> \left< \varphi_j \right| + \left| \varphi_j 
     \right> \left< \varphi_i \right| \Bigr\},
\label{eq:hamiltonian}
\end{equation}
where the $\left| \varphi_i \right>$'s are tight-binding basis 
functions centered on site $i$, $N_s$ is the total number of lattice 
sites available for the particle to hop into, and the sum in the 
second term is only over nearest-neighbor sites.  Randomly 
varying the on-site energies $\epsilon_i$ identifies the model as 
the Anderson model of localization \cite{anderson58}.  Alternatively,
setting the $\epsilon_i$ to a constant, choosing $v_{ij}$ to be a
non-zero constant only for nearest neighbor sites, and embedding the 
transport in a disordered site-percolation cluster identifies the 
model as the quantum percolation model \cite{degennes59,kirkpatrick72}.
Both the Anderson model and quantum percolation involve disorder.  
In this paper we follow the quantum percolation model, i.e., we
set $\epsilon_i\!=\!0$ and let $v_{ij}$ be a non-zero constant only 
for nearest neighbor sites.  The model, however, does not involve
disorder.  Particle transport is through perfectly ordered honeycomb
lattice ribbons.

\begin{figure}[h!]
\centering
{\includegraphics[width=3in]{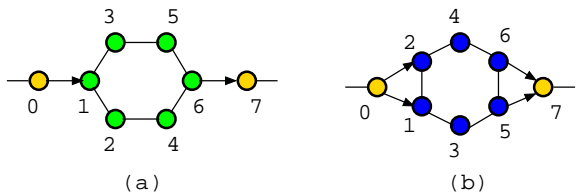}}
\caption{(color online) Hexagons in (a) armchair and (b) zigzag 
orientations.  Semi-infinite input and output linear chains are
connected to the hexagons.  Beside each site is its label.  Sites
labelled $0$ and $N_{\rm r}\!+\!1$ are the input and output supersites, 
respectively.  $N_{\rm r}$ is the number of sites within the
hexagon.
\label{fig:labels}}
\end{figure}

Shown in Fig.~\ref{fig:labels} are hexagons oriented so that they
form either armchair or zigzag edges in a honeycomb lattice 
ribbon.  In a ribbon containing $N_{\rm r}$ sites, the site with 
label $0$ in the input chain and label $N_{\rm r}\!+\!1$ in the output 
chain are the only sites that are directly connected to the ribbon.
We call these sites the input and output supersites, respectively.
The connections between these supersites and the ribbon can be 
single-channel as shown in Fig.~\ref{fig:labels}(a) or multi-channel 
as shown in Fig.~\ref{fig:labels}(b).  The distance between 
nearest-neighbor sites is set to $a_0\!=\!1$, thus setting the length 
scale of the model (although the length scale is actually 
inconsequential since $\mathcal{H}$ has no explicit length dependence).

\begin{figure}[h!]
\centering
{\includegraphics[width=3.3in,height=2.75in]{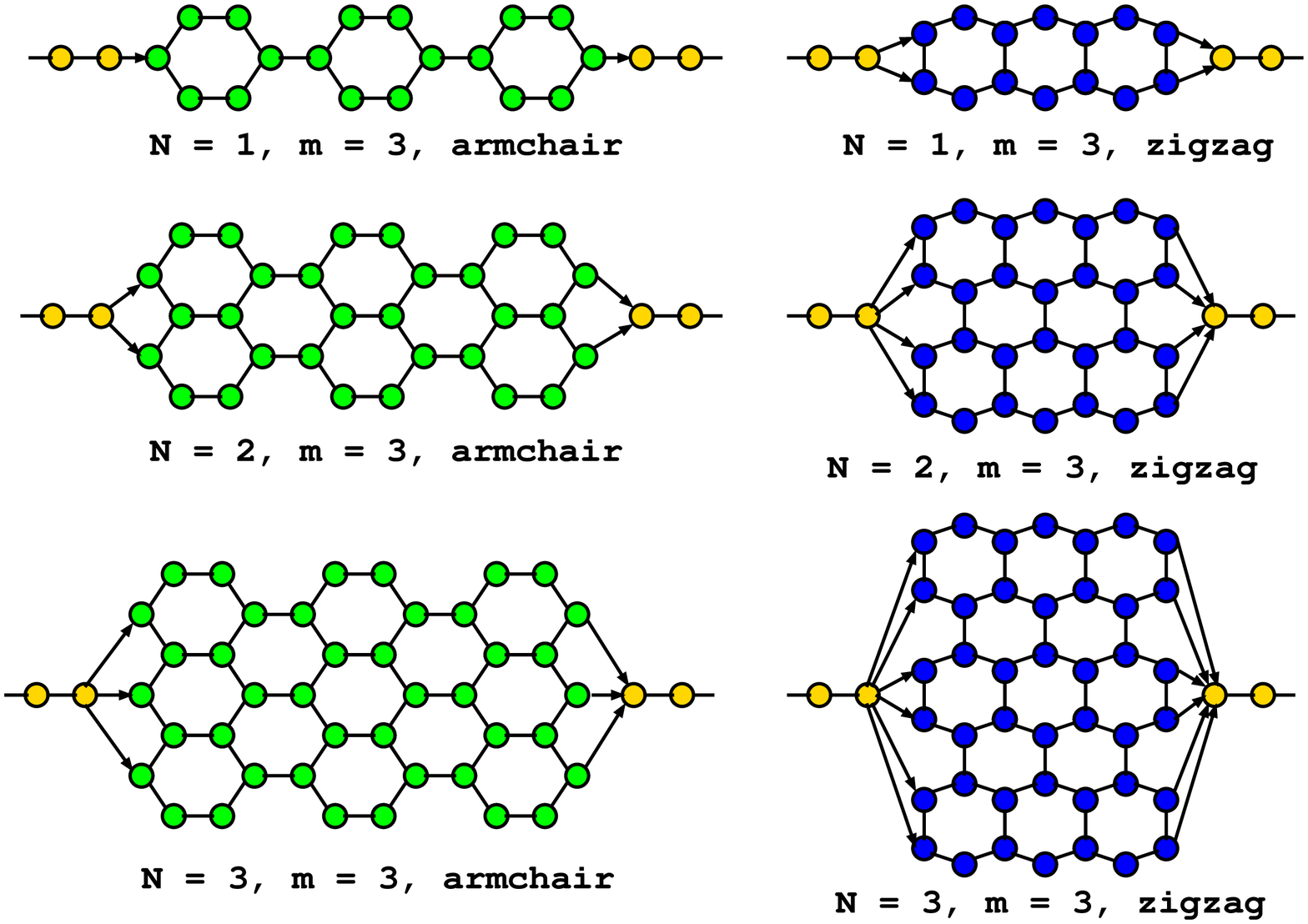}}
\caption{(color online) Honeycomb lattice ribbons of various 
widths.  $N$ is the width label of a ribbon and $m$ is the number 
of hexagons in one hexagonal chain.  Connections between the 
ribbon and the linear chain leads are also shown.
\label{fig:ribbons}}
\end{figure}

Shown in Fig.~\ref{fig:ribbons} are honeycomb lattice ribbons
with either armchair or zigzag edges and with varying widths.
The $N\!=\!1$ ribbon is a chain of hexagons coupled to the input 
and output leads.  $m$ counts the number of hexagons in one
hexagonal chain.  The $N\!=\!2$ ribbon consists of two $N\!=\!1$
ribbons placed side by side.  $N$ thus directly determines the 
ribbon width while $m$ determines the ribbon length.

We have used semi-infinite linear chains as the input and output
leads to mimic linear molecular wires.  Because of this linearity,
an incident particle having a specific energy can have only one 
mode of either a propagating or evanescent wave \cite{bagwell90} 
impinging on the ribbon.  We thus determine the effects of small 
honeycomb lattice ribbons of varying widths and lengths on the 
transmission of such single-mode waves.  Recent studies have been 
done modelling the leads as normal metals having an aggregate 
number of modes incident on large honeycomb lattice 
ribbons \cite{schomerus07,blanter07,robinson07}.  The leads were 
modelled as either square or honeycomb lattices.  It was found 
that for sufficiently large ribbons the geometrical details of 
the leads were not critical \cite{schomerus07}.  At the Dirac 
point of the ribbon evanescent waves are dominant while away from 
this point propagating waves dominate the 
transport \cite{blanter07}. In this study, in contrast, we deal 
with leads having only one mode incident on a small honeycomb 
lattice ribbon and see that the transmission of this incident wave 
depends strongly on the width, length, and edge type of the 
ribbon.

To determine the hopping transport properties of particles 
through honeycomb lattice ribbons the Hamiltonian in
Eq.~(\ref{eq:hamiltonian}) is cast in a matrix representation with
the $\left| \varphi_i \right>$'s as the basis.  We write the 
eigenvector as $\vec{\psi}$, which is a column vector with elements 
$\varphi_i$ as the coefficients of the chosen basis.  Because the 
linear chain leads are infinite the equivalent matrix problem is 
infinite-dimensional.  The contribution of the leads (except for 
the supersites), however, can be factored out since each site in 
the chain only interacts with its nearest-neighbor sites to the 
left and right, thereby resulting in predictable tridiagonal 
elements in the Hamiltonian matrix.  We follow the method proposed
by Daboul {\it et al.} \cite{daboul00} by using the scattering
boundary condition ansatz
\begin{equation}
   \varphi_n = \left\{ \begin{array}{ll}
      e^{i n q} + r\: e^{-i n q}, &~~{\rm left~input~lead},\\
      t\: e^{i(n-N_r-1)q}, &~~{\rm right~output~lead},\\
   \end{array} \right.
\label{eq:ansatz}
\end{equation}
where $r$ and $t$ are the reflection and transmission amplitudes,
respectively.  Note that along the left input lead $n\!\leq\!0$ while
along the right output lead $n\!>\!N_r$.  The ansatz states that an 
incoming plane wave and a reflected wave with amplitude $r$ are in 
the input lead and a transmitted wave with amplitude $t$ is in the 
output lead.  An alternative but equivalent approach is to use the
non-equilibrium Green's function technique to determine the
transmission \cite{wang08}.

In general the hopping amplitudes along the left and right leads and
the ribbon can be different.  In this paper we consider the left and
right leads to be the same and thus, the leads have the same hopping 
amplitudes $v_{\rm left} = v_{\rm right} = v_{\rm lead}$.  Let $v$ be 
the hopping amplitude within the ribbon.

Using the ansatz we can reduce the original infinite-dimensional 
problem into a finite one.  Along both leads we get the dispersion
relation
\begin{equation}
  e^{i q} + e^{-i q} = \frac{E}{v_{\rm lead}}.
\label{eq:leads}
\end{equation}
Choosing a numerical value for $E$ and $v_{\rm lead}$ sets the value 
of $q$.  Also, because of Eq.~\ref{eq:leads}, the value of 
$E/v_{\rm lead}$ is bounded within the interval $\left[-2,2\right]$.  
For the ribbon, the problem reduces to a finite linear problem of 
the form $\tens{A} \vec{\psi} = \vec{b}$.  In the armchair hexagon 
shown in Fig.~\ref{fig:labels}(a), for example, we get
\begin{displaymath}
   \tens{A} = \left( \begin{array}{cccccccc}
      -\alpha\!+\! i \beta & v_{\rm lead} & 0 & 0 & 0 & 0 & 0 &   0 \\
          v_{\rm lead}    & -E & v & v & 0 & 0 & 0 &   0 \\
          0    & v & -E & 0 & v & 0 & 0 &   0 \\
          0    & v & 0 & -E & 0 & v & 0 &   0 \\
          0    & 0 & v & 0 & -E & 0 & v &   0 \\
          0    & 0 & 0 & v & 0 & -E & v &   0 \\
          0    & 0 & 0 & 0 & v & v & -E &   v_{\rm lead} \\
          0    & 0 & 0 & 0 & 0 & 0 & v_{\rm lead} & -\alpha\!+\! i \beta \\
   \end{array} \right),
\end{displaymath}
\begin{equation}
   \vec{\psi} = \left( \begin{array}{c}
       1\!+\!r \\ \varphi_1 \\ \varphi_2 \\ \varphi_3 \\
       \varphi_4 \\ \varphi_5 \\ \varphi_6 \\ t \\
   \end{array} \right),\ {\rm and}\ 
   \vec{b} = \left( \begin{array}{c}
      2 i \beta \\ 0 \\ 0 \\ 0 \\ 0 \\ 0 \\ 0 \\ 0 \\
   \end{array} \right),
\label{eq:linear}
\end{equation}
where $\alpha\!=\!E/2$, $\beta\!=\!\sqrt{4 v^2_{\rm lead}\!-\!E^2}/2$.

We can determine $\vec{\psi}$ exactly by carefully taking the 
inverse of $\tens{A}$ and then multiplying it to $\vec{b}$.  To do 
this we numerically decompose $\tens{A}$ using singular-value 
decomposition so that 
$\tens{A} = \tens{U}\:\tens{\Sigma}\:\tens{V}^{\rm \dagger}$, 
where $\tens{U}$ and $\tens{V}$ are unitary matrices and
$\tens{\Sigma}$ is a diagonal matrix \cite{golub96}.  The inverse 
can then be determined using the properties of unitary and 
diagonal matrices.  Once we determine $\vec{\psi}$ we can then 
calculate the transmission coefficient by $T\!=\!\left| t \right|^2$.  
This numerically exact approach has also been employed in the 
study of localization in disordered clusters in two-dimensional 
quantum percolation \cite{cuansing08,islam08}.

The connections between the input and output supersites and the
ribbon are reflected in the outermost rows and columns of 
$\tens{A}$.  The inner rows and columns reflect the connections
between sites within the ribbon.  And so for the zigzag hexagon 
shown in Fig.~\ref{fig:labels}(b) the corresponding matrix $\tens{A}$ 
will have almost the same central entries as that shown in 
Eq.~(\ref{eq:linear}) except for the extra entries along the outer
rows and columns because of the additional channels available 
between the supersites and the hexagon.

A merit of determining all elements of $\psi$ exactly is that the 
components $\varphi_i$ of the wave function at all of the sites in 
the ribbon are also determined.  In Sec.~\ref{sec:results} we show 
plots of the wave function density in $N\!=\!1$, $m\!=\!8$ ribbons with 
armchair or zigzag edges.

\section{Numerical results \label{sec:results}}

Depending on the atomic composition of the ribbon and the leads, $v$
and $v_{\rm lead}$ are in general not the same.  In our simulations,
however, we set the values of $v$ and $v_{\rm lead}$ to one.  We vary 
the energy from $E\!=\!-2$ to $E\!=\!2$ in energy steps of 
$\Delta E\!=\!0.01$.  For each value of $E$ the complex-valued matrix 
$\tens{A}$ is constructed by noting the connections between the 
sites, including the supersites.  The function $\vec{\psi}$ is then 
numerically determined via the singular-value decomposition of 
$\tens{A}$.  The transmission coefficient $T$ is then calculated from 
$\left| t \right|^2$.

\begin{figure}[h!]
\centering
{\includegraphics[width=2.65in]{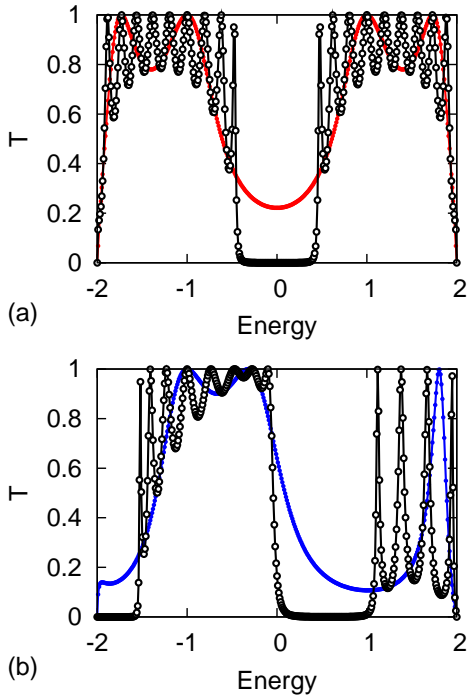}}
\caption{(color online) Transmission coefficient $T$ as a function 
of the particle's energy in honeycomb lattice ribbons of width 
$N\!=\!1$. (a) Armchair edges with  $m\!=\!2$ for filled (red online) 
circles and $m\!=\!8$ for open circles.  The lines drawn are not 
analytical fits but simply connect adjacent datapoints to aid the 
eye. (b) Zigzag edges with $m\!=\!2$ for filled (blue online) 
circles and $m\!=\!8$ for open circles.
\label{fig:length}}
\end{figure}

Shown in Fig.~\ref{fig:length} are plots of the transmission 
coefficient $T$ as a function of the energy $E$ of the particle in 
$N\!=\!1$ ribbons.  Ribbons with armchair edges are shown in 
Fig.~\ref{fig:length}(a) while those with zigzag edges are shown in 
Fig.~\ref{fig:length}(b).  The length of the ribbons are $m\!=\!2$
and $m\!=\!8$ hexagons.  For the ribbons with armchair edges we see 
symmetry between the positive and negative energy regions, i.e., with 
respect to the middle of the band.  This symmetry in the sign of the 
energy arises from a condition satisfied by the $\tens{A}$ matrix for 
the ribbon with armchair edges and is discussed in 
Sec.~\ref{sec:symmetry}.  We also see two features as the length of the 
ribbons are increased from $m\!=\!2$ (red dots) to $m\!=\!8$ (black
dots): (a) the number of transmission oscillations increases 
proportionally, and (b) a transmission gap develops.  The transmission 
oscillations are reminiscent of the transmission resonances found in 
the square lattice case \cite{cuansing04}.  The transmission gap 
develops as the length of the ribbon is extended.  Although $T$ has a 
non-zero minimum at the middle of the band for $m\!=\!2$, this 
minimum falls to zero transmission for the first time when $m\!=\!6$.  
Extending the ribbon length further to $m\!=\!100$ increases the 
number of transmission oscillations and sharpens, but does not change 
the width, of the gap.

For the ribbons with zigzag edges we see in Fig.~\ref{fig:length}(b)
that the number of transmission oscillations grows and a 
transmission gap develops as the length of the ribbon is extended 
from $m\!=\!2$ (blue dots) to $m\!=\!8$ (black dots), similar to 
the case for armchair edges.  However, the center of the transmission 
gap is not at the middle of the band, which is now slightly 
transmitting.  And in contrast to the armchair edges case, the number 
of oscillations in the $E\!>\!0$ region is half those in the 
$E\!<\!0$ region.  The absence of symmetry with respect to the middle 
of the band in the transmission for ribbons with zigzag edges can be 
attributed to the extra connections between the input and output 
supersites and the ribbon, as discussed in Sec.~\ref{sec:symmetry}.

\begin{figure}[h!]
\centering
{\includegraphics[width=3.4in]{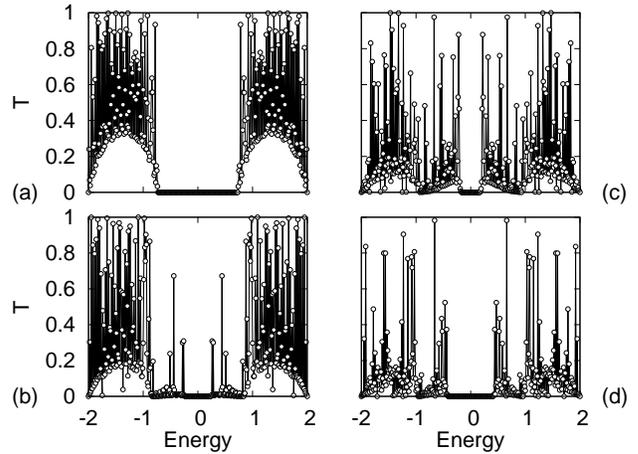}}
\caption{Transmission coefficient $T$ as a function of the 
particle's energy $E$ in honeycomb lattice ribbons with armchair
edges.  The ribbons have length $m\!=\!100$.  The widths are 
(a) $N\!=\!2$, (b) $N\!=\!3$, (c) $N\!=\!4$, and (d) $N\!=\!5$.
The transmission gap is symmetric with respect to the middle of
the band.
\label{fig:armchair100}}
\end{figure}

\begin{figure}[h!]
\centering
{\includegraphics[width=3.4in]{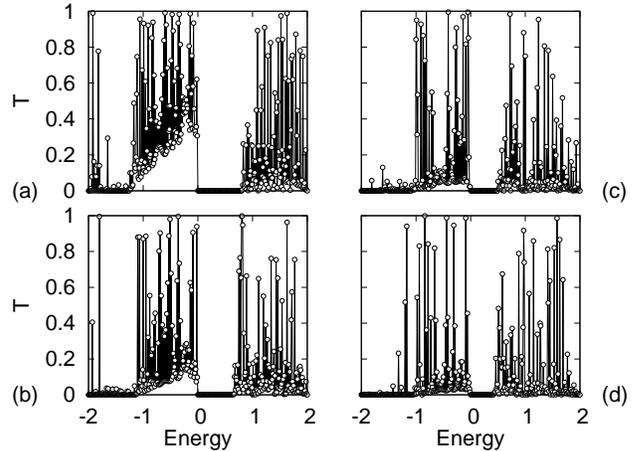}}
\caption{Transmission coefficient $T$ as a function of the
particle's energy $E$ in honeycomb lattice ribbons with zigzag
edges.  The ribbons have length $m\!=\!100$.  The widths are
(a) $N\!=\!2$, (b) $N\!=\!3$, (c) $N\!=\!4$, and (d) $N\!=\!5$.
The transmission gap shrinks systematically as the ribbon's
width is increased.
\label{fig:zigzag100}}
\end{figure}

Shown in Figs.~\ref{fig:armchair100} and \ref{fig:zigzag100} are 
plots on how the transmission coefficient behaves as a function of 
the energy of the particle as the width of the ribbons is varied from 
$N\!=\!2$ to $N\!=\!5$.  The ribbons have length $m\!=\!100$ and have 
armchair edges in Fig.~\ref{fig:armchair100} and zigzag edges in 
Fig.~\ref{fig:zigzag100}.  For the ribbons with armchair edges the 
transmission gap is symmetric with respect to the middle of the band.
In contrast, for the ribbons with zigzag edges the center of the 
transmisison gap does not coincide with the middle of the band.  In 
addition, the width of the gap decreases systematically as the width 
of the ribbon with zigzag edges is increased.  This is not the case 
for the ribbons with armchair edges.  Taking a closer look at
Fig.~\ref{fig:armchair100}(b) we see that transmission resonances
appear in the region where there is a gap in the narrower ribbon
shown in Fig.~\ref{fig:armchair100}(a).  As the ribbon's width is 
increased transmission resonances appear within the gap and thus
obscures the determination of the exact width of the gap.

We thus see in narrow ribbons with either armchair or zigzag edges 
a transmission gap dividing the transmission curve into two 
regions.  As the width of the ribbon is increased the gap in ribbons 
with armchair edges is obscured by the appearance of transmission 
resonances within it while the gap in ribbons with zigzag edges 
systematically shrinks.

\begin{figure}[h!]
\centering
{\includegraphics[width=3in]{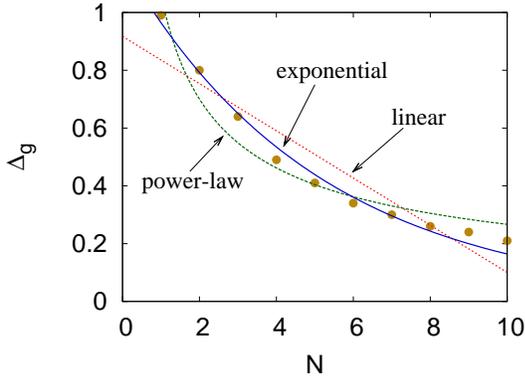}}
\caption{(color online) Transmission gap, $\Delta_g$, as a 
function of the width $N$ for $m\!=\!100$ ribbons with zigzag 
edges.  Linear (red), power-law (green) and exponential (blue) 
fitting curves are also shown.
\label{fig:fit}}
\end{figure}

The dependence of the transmission gap, $\Delta_g$, on the width
of the ribbon with zigzag edges can be determined.  For
ribbons with armchair edges, in contrast, the gap is not 
well-defined because of the appearance of resonances and thus 
there is no clear indication of the gap's width.  Shown in 
Fig.~\ref{fig:fit} is a plot of $\Delta_g$ as a function of $N$ for
ribbons with zigzag edges.  The length of the ribbons are $m\!=\!100$.  
Also shown are the curves for linear, power-law and exponential fits.
The values of the fitting parameters are shown in Table~\ref{tab:fits}.
The correlation coefficients found for the three fits are close, with
the exponential fit having the best value.  

\begin{table}
\caption{Values of the fitting parameters found for the data shown
in Fig.~\ref{fig:fit} for ribbons with zigzag edges.
\label{tab:fits}}
\begin{tabular}{l}
  \hline\hline\noalign{\smallskip}
  linear fit: \\
  $\Delta_g = (-0.082 \pm 0.010)\:N + (0.917 \pm 0.065)$,\  
    $\left| R \right|^2 = 0.928$ \\ \\
  power-law fit: \\
  $\Delta_g = (1.061 \pm 0.058)\:N^{(-0.600 \pm 0.049)}$,\ 
    $\left| R \right|^2 = 0.967$ \\ \\
  exponential fit: \\
  $\Delta_g = (1.175 \pm 0.038)\:e^{(-0.197 \pm 0.009)\:N}$,\ 
    $\left| R \right|^2 = 0.992$ \\
  \noalign{\smallskip}\hline\hline
\end{tabular}
\end{table}

\begin{figure}[h!]
\centering
{\includegraphics[width=3.3in]{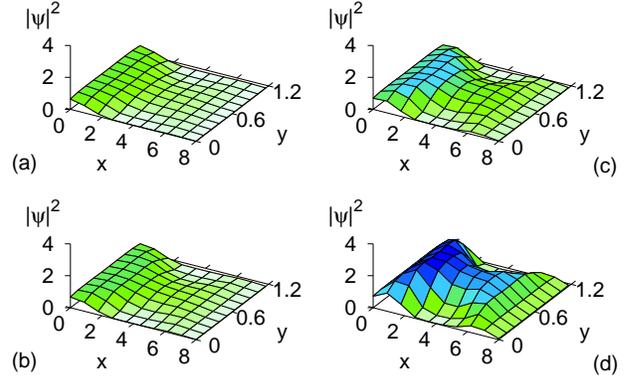}}
\caption{(color online) Wave function density in $N\!=\!1$,
$m\!=\!8$ ribbons with zigzag edges.  The energies are 
(a) $E\!=\!1.03$, (b) $E\!=\!1.05$, (c) $E\!=\!1.07$, and 
(d) $E\!=\!1.09$.  These energies are in the region around the 
right edge of the transmission gap.  The colors aid the eye in 
determining the height of $\left| \vec{\psi} \right|^2$.
\label{fig:gap}}
\end{figure}

Shown in Fig.~\ref{fig:gap} is the evolution of the wavefunction 
density $\left| \vec{\psi} \right|^2$ as the energy $E$ is varied 
from within the transmission gap into the $E\!>\!0$ transmitting 
region of the $N\!=\!1$, $m\!=\!8$ ribbon with zigzag edges.  In the 
figure, the ribbon is along the $xy$-plane.  The input lead is 
coupled to the ribbon at $x\!=\!0$ and the output lead is coupled at 
the other side.  For energies within the transmission gap the wave 
function is zero.  For energies near the edge of the gap, for
example at $E\!=\!1.03$ as shown in Fig.~\ref{fig:gap}(a), the wave 
function is almost zero everywhere except for a slight hump at 
the input side of the ribbon.  This hump grows as the energy is 
increased until within the transmitting region a well-developed 
wave with a relatively large amplitude appears.  Within the gap the 
wave is evanescent \cite{bagwell90} and trying to penetrate the 
ribbon.  As we increase the energy of the particle the wave 
eventually gains enough energy to proceed into the ribbon and 
transmit to the other side.

\begin{figure}[h!]
\centering
{\includegraphics[width=3.3in]{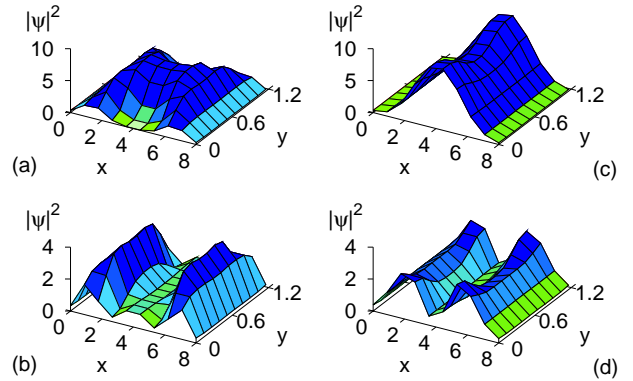}}
\caption{(color online) The $\left| \vec{\psi} \right|^2$ in 
$N\!=\!1$, $m\!=\!8$ ribbons with zigzag edges.  These waves 
correspond to transmission peaks with energy values (a) $E\!=\!1.12$, 
(b) $E\!=\!1.38$, (c) $E\!=\!-1.52$, and (d) $E\!=\!-1.42$.
\label{fig:zigpeaks}}
\end{figure}

Shown in Fig.~\ref{fig:zigpeaks} are some of the wave function
densities that correspond to transmission peaks in $N\!=\!1$, 
$m\!=\!8$ ribbons with zigzag edges.  The wave in 
Fig.~\ref{fig:zigpeaks}(a) is the highly transmitting wave that the 
waves in Fig.~\ref{fig:gap} develop into as the particle's energy 
is increased.  Compared to the waves in Fig.~\ref{fig:gap} the 
highly transmitting wave in Fig.~\ref{fig:zigpeaks}(a) is sitting 
symmetrically at the center of the ribbon.  This characteristic is 
a property satisfied by the wave functions that correspond to the
transmission peaks in Fig.~\ref{fig:length}, including those for the
armchair edges.

\section{Symmetry of ribbons with armchair edges \label{sec:symmetry}}

In Fig.~\ref{fig:length} the transmission coefficient for ribbons 
with armchair edges is shown to be symmetric with respect to a 
change in the sign of the energy.  This symmetry arises from a 
property of the $\tens{A}$ matrix for the ribbon.  

Consider, for example, the linear problem 
$\tens{A} \vec{\psi}\!=\!\vec{b}$ shown in Eq.~(\ref{eq:linear}) for 
the armchair hexagon.  For a ribbon of $N_r$ sites the corresponding 
matrix $\tens{A}$ has size 
$\left(N_r\!+\!2\right) \times \left(N_r\!+\!2\right)$ and 
$\vec{\psi}$ and $\vec{b}$ are column vectors of size 
$\left(N_r\!+\!2\right)$.  Suppose we have two linear problems, 
$\tens{A} \vec{\psi}_1 = \vec{b}$ and 
$\tens{B} \vec{\psi}_2 = \vec{b}$, with the same right-hand side 
$\vec{b}$.  The $\vec{\psi}_1$ and $\vec{\psi}_2$ can be determined 
using the inverses of $\tens{A}$ and $\tens{B}$, i.e., 
$\vec{\psi}_1 = \tens{A}^{-1} \vec{b}$ and 
$\vec{\psi}_2 = \tens{B}^{-1} \vec{b}$.  We can write the inverses of 
the matrices in terms of their corresponding matrix of cofactors, 
\begin{equation}
  \tens{A} = \frac{\tens{C}^{\rm{T}}}{\left|\tens{A}\right|},
  \ \rm{and}\
  \tens{B} = \frac{\tens{D}^{\rm{T}}}{\left|\tens{B}\right|},
\end{equation}
where $\left|\tens{A}\right|$ and $\left|\tens{B}\right|$ are
the determinants of the matrices.  Since $\tens{A}$ and 
$\tens{B}$ are symmetric, $\tens{C}^{\rm{T}} = \tens{C}$ and
$\tens{D}^{\rm{T}} = \tens{D}$.  In particular, $\vec{\psi}_1$ can be
calculated from
\begin{equation}
\left( \begin{array}{c}
  1\!+\!r_1 \\ \varphi_1 \\ \vdots \\ \varphi_N \\ t_1
\end{array} \right)
= \frac{1}{\left| \tens{A} \right|}
\left( \begin{array}{ccc}
  C_{0,0} & \cdots & C_{0,N_r\!+\!1} \\
  C_{1,0} & \cdots & C_{1,N_r\!+\!1} \\
  \vdots & \ddots & \vdots \\
  C_{N_r,0} & \cdots & C_{N_r,N_r} \\
  C_{N_r\!+\!1,0} & \cdots & C_{N_r\!+\!1,N_r\!+\!1} \\
\end{array} \right)
\left( \begin{array}{c}
  2 i \beta \\ 0 \\ \vdots \\ 0 \\ 0 \\
\end{array} \right).
\label{eq:x1}
\end{equation}
The transmission amplitude $t_1$ can be determined from the
multiplication of the last row of $\tens{C}$ with $\vec{b}$.  We get
\begin{equation}
t_1 = \frac{2\,i\,\beta\,C_{N_r\!+\!1,0}}{\left| \tens{A} \right|}.
\label{eq:t1}
\end{equation}
A similar calculation can be done to determine $t_2$.  The
transmission coefficients can then be calculated from
$T_1 = \left| t_1 \right|^2$ and $T_2 = \left| t_2 \right|^2$.

Suppose $\tens{A}\!=\!\tens{A}(E)$ and $\tens{B}\!=\!\tens{A}(-E)$.  
The condition that a change in the sign of $E$ results in the same 
transmission coefficient, using Eq.~(\ref{eq:t1}), is
\begin{equation}
\left| C(E)_{N_r\!+\!1,0} \right|^2 = 
\left| C(-E)_{N_r\!+\!1,0} \right|^2,
\label{eq:condition}
\end{equation}
where we use the fact that 
$\left| \tens{A}(E) \right|^2 = \left| \tens{A}(-E) \right|^2$.
For there to be symmetry between the positive and negative energy
regions, Eq.~(\ref{eq:condition}) must be satisfied.  Note that this
condition is independent of the size of the ribbon, as long as that
ribbon is coupled to input and output leads.

Calculating the cofactor $C(E)_{7,0}$ of $\tens{A}$ in 
Eq.~(\ref{eq:linear}), noting that in our simulations $v_l = v$, 
results in
\begin{equation}
\left| C(E)_{7,0} \right|^2 = 4 v^{10} \left( E^4 - 2 v^2 E^2 + v^4
\right).
\label{eq:armcofact}
\end{equation}
This is an even function of $E$ and thus satisfies the symmetry 
condition of Eq.~(\ref{eq:condition}).  In contrast, for the zigzag 
hexagon its corresponding $\tens{A}$ matrix will have four more $v$ 
entries than the one shown in Eq.~(\ref{eq:linear}).  This reflects 
the second channel available in the coupling between the supersites 
and the hexagon.  The cofactor of the resulting $\tens{A}$ for this 
case leads to
\begin{equation}
\begin{array}{rcr}
\left| C(E)_{7,0} \right|^2 & = & 4 v^8 (E^6 + 4 v E^5 + 2 v^2 E^4
  - 8 v^3 E^3 \\
  & & - 7 v^4 E^2 + 4 v^5 E + 4 v^6). \\
\end{array}
\label{eq:zigcofact}
\end{equation}
Terms with odd powers of $E$ appear and the cofactor therefore will
be different when the sign of $E$ is reversed.  The symmetry
condition in Eq.~(\ref{eq:condition}) is not satisfied and, as can
be seen in Fig.~\ref{fig:length}(b), the profile of the transmission
coefficient is different between the positive and negative energy 
regions.  This analysis can be extended to ribbons with any $N$ and
$m$.  As can be seen in Figs.~\ref{fig:armchair100} and
\ref{fig:zigzag100} the symmetry is preserved only in ribbons with 
armchair edges.

\section{Summary \label{sec:summary}}

We investigate the transport of a quantum particle traversing through 
honeycomb lattice ribbons with either armchair or zigzag edges with
varying length and width.  The ribbons are coupled to semi-infinite
linear chains that serve as the input and output leads.  We have seen
the role of coupling the leads to the ribbons.  In narrow, $N\!=\!1$, 
ribbons we find transmission gaps for both types of edges.  In 
armchair edges the center of the gap coincides with the middle of the 
band.  This symmetry is due to a property of the corresponding 
$\mathcal{A}$ matrices.  In zigzag edges the gap center is displaced 
to the positive energy region.  We also find transmission resonances 
for both types of edges.  These transmission resonances occur 
whenever the wave fits symmetrically within the ribbon, reminiscent 
of resonant tunneling but without having finite potentials or 
tunneling in this model.  Increasing the length of the ribbons 
increases the number of transmission resonances without affecting the 
width of the transmission gap, as long as the ribbon is longer than a 
critical length when the gap can form.  Increasing the width of the 
ribbons affects the width of the transmission gap.  In ribbons with 
armchair edges transmission resonances appear within the gap as the
ribbon width is extended. In ribbons with zigzag edges the gap 
systematically shrinks as the ribbon width is increased.

\begin{acknowledgement}
We would like to thank Barbaros \"{O}zyilmaz, Li-Fa Zhang and Jing-Hua 
Lan for insightful discussions.  This work is supported in part by 
a Faculty Research Grant no. R-144-000-173-101/112 from the National 
University of Singapore.
\end{acknowledgement}

\bibliographystyle{epj}
\bibliography{references}

\end{document}